\pdfoutput=1
\PassOptionsToPackage{dvipsnames}{xcolor}

\documentclass[lettersize,journal]{IEEEtran}

\IEEEoverridecommandlockouts
\usepackage{amsmath,amsfonts}
\usepackage[ruled,linesnumbered]{algorithm2e}
\usepackage{amssymb}
\usepackage{algorithmic}
\usepackage{bibentry}
\usepackage{amsfonts}
\usepackage{subfigure}
\usepackage{multirow}
\usepackage{amsmath}
\usepackage{graphicx}
\usepackage{epsfig}
\usepackage{color}
\usepackage{epstopdf}
\usepackage{ragged2e}
\usepackage{rotating}
\usepackage{caption}
\usepackage{float}
\usepackage{multicol}
\usepackage{amssymb}
\usepackage{graphicx}
\usepackage{bbding}
\usepackage{balance}
\usepackage{microtype}
\usepackage{threeparttable}
\usepackage{subfigure}
\usepackage{booktabs} 
\usepackage{multirow}
\usepackage{amsmath}
\usepackage[many]{tcolorbox}
\usepackage{mathtools}
\usepackage{etoolbox}
\usepackage{cases}
\usepackage{enumitem}

\usepackage[dvipsnames]{xcolor}
\usepackage{subfigure}
\usepackage{tabularx}
\usepackage{makecell}
\usepackage{bm}
\usepackage{colortbl}
\usepackage{bbold}

\newcommand{\FAIRNESS}{{Stereotype-aware Fairness}}
\newcommand{\fairness}{{stereotype-aware fairness}}
\newcommand{\llmrec}{{LLM-RS}}
\newcommand{\MOS}{{Mixture-of-Stereotypes (MoS)}}
\newcommand{\mos}{{MoS}}

\begin{document}

\title{Investigating and Mitigating Stereotype-aware Unfairness in LLM-based Recommendations}

\author{Zihuai Zhao, Wenqi Fan$^*$, Yao Wu, Qing Li,~\IEEEmembership{Fellow,~IEEE}
\IEEEcompsocitemizethanks{
\IEEEcompsocthanksitem Z. Zhao, and Q. Li are with the Department of Computing, The Hong Kong Polytechnic University. E-mail: zihuai.zhao@connect.polyu.hk, qing-prof.li@polyu.edu.hk.
\IEEEcompsocthanksitem W. Fan is with the Department of Computing (COMP) and Department of Management and Marketing (MM), The Hong Kong Polytechnic University. E-mail: wenqifan03@gmail.com.
\IEEEcompsocthanksitem Y. Wu is with the Department of Management and Marketing (MM), The Hong Kong Polytechnic University. E-mail: wuyao.wu@polyu.edu.hk.
}
\thanks{*Corresponding author: Wenqi Fan.}}

\markboth{IEEE TRANSACTIONS ON KNOWLEDGE AND DATA ENGINEERING, SUBMISSION 2025}%
{Shell \MakeLowercase{\textit{et al.}}: A Sample Article Using IEEEtran.cls for IEEE Journals}


\maketitle

\begin{abstract}
    Large Language Models (LLMs) have demonstrated unprecedented language understanding and reasoning capabilities to capture diverse user preferences and advance personalized recommendations.
    Despite the growing interest in LLM-based recommendations, unique challenges are brought to the trustworthiness of LLM-based recommender systems (\llmrec{}).
    Compared to unique user/item representations in conventional recommender systems, users and items share the textual representation (e.g., word embeddings) in LLM-based recommendations.
    Recent studies have revealed that LLMs are likely to inherit stereotypes that are embedded ubiquitously in word embeddings, due to their training on large-scale uncurated datasets.
    This leads to \llmrec{} exhibiting stereotypical linguistic associations between users and items, causing a form of two-sided (i.e., user-to-item) recommendation fairness.
    However, there remains a lack of studies investigating the unfairness of \llmrec{} due to intrinsic stereotypes, which can simultaneously involve user and item groups.
    To bridge this gap, this study reveals a new variant of fairness between stereotype groups containing both users and items, to quantify discrimination against stereotypes in \llmrec{}. 
    Moreover, in this paper, to mitigate stereotype-aware unfairness in textual user and item representations, we propose a novel framework named \textbf{\MOS{}}. In particular, an insightful stereotype-wise routing strategy over multiple stereotype-relevant experts is designed, aiming to learn unbiased representations against different stereotypes in LLM-RS.
    Extensive experiments are conducted to analyze the influence of \fairness{} in \llmrec{} and the effectiveness of our proposed methods, which consistently outperform competitive benchmarks under various fairness settings.
\end{abstract}


\section{INTRODUCTION}
\label{sec:introduction}

\begin{figure}[htbp]
\centering
{\includegraphics[width=\linewidth]{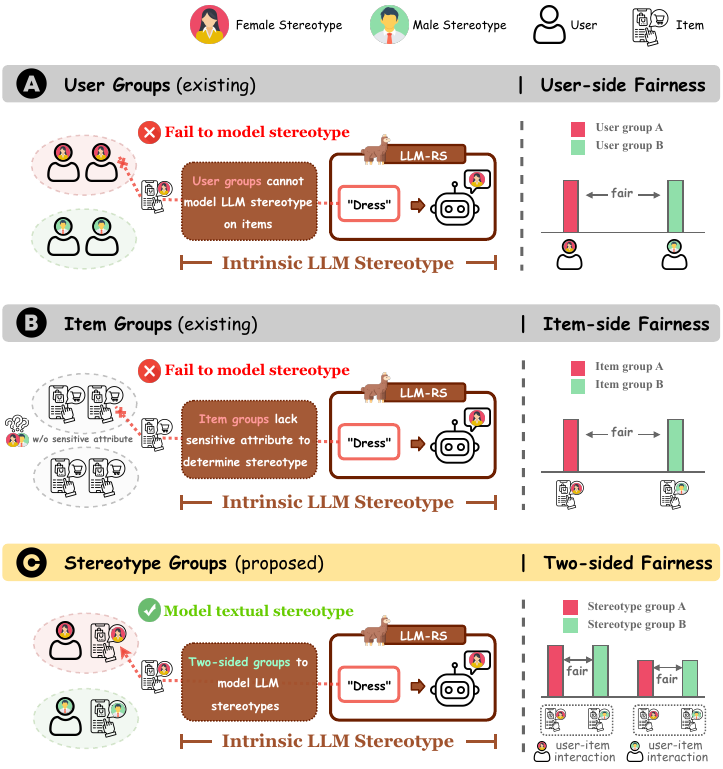}}
\caption{Illustration of stereotype groups. Compared to unique user/item representations in conventional recommender systems, users and items share the textual representation (e.g., word embeddings) in LLM-based recommendations. Therefore, LLM stereotypes can simultaneously involve user and item groups, leading to a form of two-sided recommendation fairness.}
\vskip -0.15in
\label{fig:group}
\end{figure}

\IEEEPARstart{R}{ECOMMENDER} systems (RS) provide personalized suggestions tailored to user preferences, facilitating user experience across diverse applications, such as e-commerce~\cite{jin2024amazon}, job matching~\cite{wu2024exploring}, and social media platforms~\cite{fan2019graph, fan2020graph}.
Recently, Large Language Models (LLMs) have emerged as a prevalent paradigm for advancing personalized recommendations.
To be specific, LLMs equipped with billion-scale parameters have demonstrated unprecedented language understanding and reasoning capabilities to capture diverse user preferences based on rich textual side information in RS (e.g., user profiles and item descriptions)~\cite{zhao2024recommender}.
However, the integration of LLMs into recommendations brings about unique challenges toward the trustworthiness of LLM-based recommender systems (\textbf{\llmrec{}}). Specifically, recent studies have revealed that LLMs trained on large-scale uncurated data inherit stereotypes against social groups, leading to intrinsic biases in downstream applications~\cite{navigli2023biases, gallegos2024bias}.
For instance, LLMs tend to overlook the personalized preference behind user-item interactions but simply perform recommendations based on stereotypical textual knowledge, such as suggesting ``\emph{female nurse}" and ``\emph{male doctor}" in job recommendations~\cite{gallegos2024bias, kotek2023gender}.

\begin{figure*}[t]
    \centering
    \includegraphics[width=\textwidth]{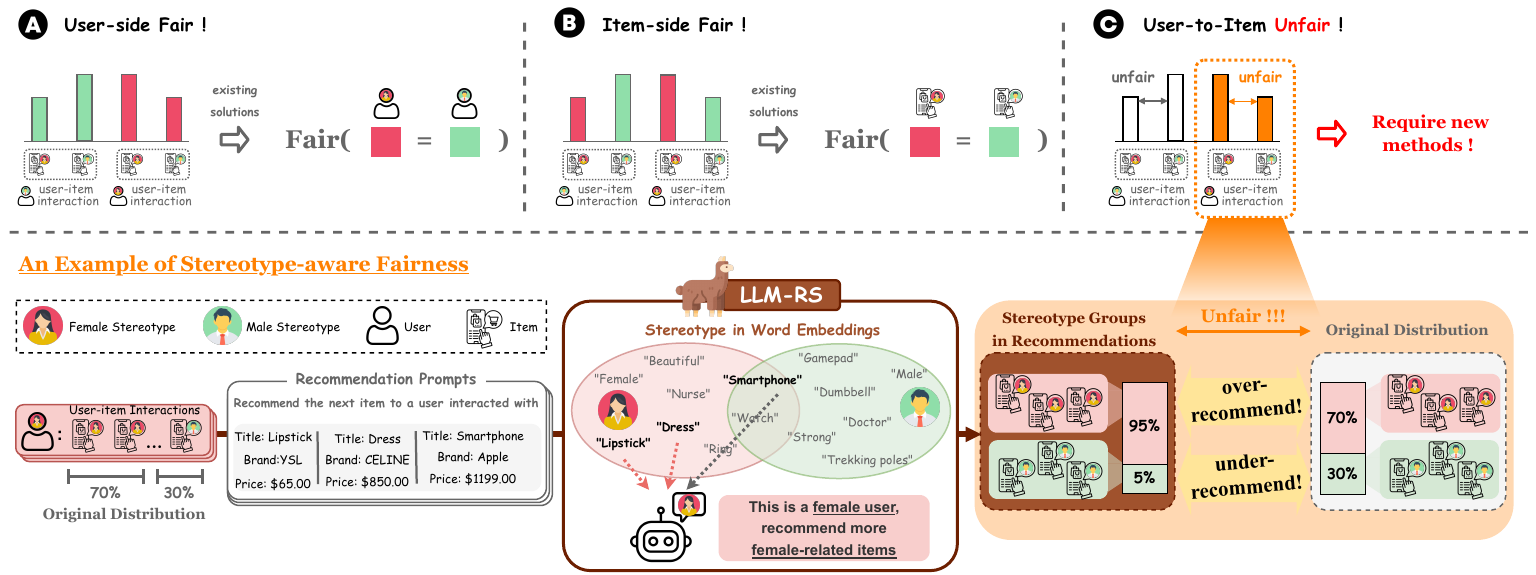}
    \caption{Illustration of stereotype-aware fairness. In A and B, existing user-side or item-side fairness still ignores two-sided (i.e., user-to-item) stereotypes, only considering the fairness against single-sided stereotypes.
    In C, we model the fairness against two-sided (i.e., user-to-item) stereotypes based on \textbf{calibrated proportions}. For example, the user-side stereotype can unfairly amplify the original distribution of item-side stereotypes, by over-recommending items (i.e., increase from 70\% to 95\%) in the female stereotype group and under-recommending items (i.e., decrease from 30\% to 5\%) in the male stereotype group.}
\label{fig:fairness}  
\end{figure*}

The fairness in recommendations can be categorized into three types according to the stakeholders involved in modeling user-item interactions, namely user-side, item-side, and two-sided fairness~\cite{wang2023survey, chen2023bias}.
Most existing studies on \llmrec{} fairness focus on either user-side fairness to achieve consistent recommendation performance across user groups~\cite{hua2023up5, zhang2023chatgpt, deldjoo2024cfairllm} or item-side fairness by providing fair exposure opportunities across item groups~\cite{jiang2024item, bao2024large}.
However, as shown in Figure~\ref{fig:group}, LLM-encoded stereotypes can simultaneously involve user and item groups, leading to a form of two-sided (i.e., user-to-item) recommendation fairness.
This is because stereotypes are embedded ubiquitously in the word embeddings of LLMs~\cite{navigli2023biases,gallegos2024bias}, where users and items share these representations (e.g., word embeddings) in LLM-based recommendations. 
For example, the same vocabulary of word embeddings is used by LLMs to represent user profiles and item descriptions.
This leads to \llmrec{} exhibiting unfair recommendations due to stereotypical linguistic associations between users and items, as illustrated in Figure~\ref{fig:fairness}, we take the female stereotype as an example.
Due to stereotypes in word embeddings, \llmrec{} tend to group a user's historical interactions into the female stereotype.
As a result, the user-side stereotype can unfairly amplify the original distribution of item-side stereotypes, by over-recommending items (i.e., increase from 70\% to 95\%) in the female stereotype group and under-recommending items (i.e., decrease from 30\% to 5\%) in the male stereotype group.

Failing to address the unfairness due to stereotypes in LLM-based recommendations can significantly reduce user satisfaction and diversity of personalized recommendations, since stereotypes might hinder \llmrec{} from exploring potential items for users.
Moreover, harmful stereotypes of LLMs could further reinforce social polarization in recommendations, such as suggesting low-paid jobs to certain gender identities and nationalities~\cite{wang2024intersectional}.
Therefore, it is imperative to delve into the fairness against stereotypes in LLM-based recommendations.

Due to the simultaneous existence of stereotypes in user and item groups, as explained in Fig.~\ref{fig:group}, existing user-side or item-side fairness could fall short of modeling textual stereotypes.
Therefore, we propose a new variant of fairness between stereotype groups containing both users and items (i.e., two-sided groups), rather than separating user groups and item groups.
To validate the existence of stereotypes and quantify the recommendation unfairness between different stereotype groups, we design a new evaluation metric named \textbf{\fairness{}} and conduct preliminary experiments on real-world recommendation datasets. As detailed in Section~\ref{sec:analysis}, our findings demonstrate that \llmrec{} can exhibit significant discrimination between different stereotype groups, highlighting the concern of \fairness{} toward the
trustworthiness of LLM-based recommendations.

To mitigate unfairness caused by stereotypes, unique challenges are brought to LLM-based recommendations. Compared to conventional recommender systems, users and items are \textbf{not} consolidated into unique representations, but span over a sequence of textual representations (e.g., item descriptions) in \llmrec{}~\cite{qu2024tokenrec, hua2023index}.
For example, most existing fairness criteria in RS, which are used to measure the similarity between user and item embeddings, are inapplicable to word sequences~\cite{wang2024intersectional, wan2020addressing}. This leads to significant difficulties in distinguishing different stereotype groups based on the textual information of users and items in \llmrec{}. 
Recently, studies have revealed that LLMs possess virtual personalities that are sensitive to prompt biases~\cite{rao2023can, xu2023llms}.
For example, LLM agents can exhibit diverse human-like personalities by giving user profiles in prompts~\cite{frisch2024llm, la2024open}, uncovering the great potential to assign different stereotype roles to LLMs.
Building up these insights, we propose a novel method named \textbf{\MOS{}} to capture and mitigate different stereotypes in \llmrec{}, utilizing a set of stereotype-relevant experts (i.e., multiple stereotype roles).
More specifically, we develop an insightful routing strategy over multiple stereotype-relevant experts to learn unbiased representations, which can be seamlessly integrated with LLMs via parameter-efficient fine-tuning (PEFT) like prompt tuning, as existing research has demonstrated the effectiveness of adapting multi-expert structures to the training of LLMs via PEFT paradigms~\cite{zadouri2023pushing, choi2023smop}. 

The main contributions of this paper are summarized as follows: 
\begin{itemize}[leftmargin=*]
    \item 
    This study investigates the unique characteristics of LLM stereotypes in recommendation unfairness, caused by the stereotypical linguistic associations between users and items.
    In this paper, we propose a new variant of fairness against stereotypes, which simultaneously involves user and item groups (i.e., two-sided groups) in \llmrec{}. 
    
    \item We propose a novel framework, namely \textbf{\MOS{}}, to mitigate discrimination against stereotypes in LLM-based recommendations. Notably, an insightful stereotype-wise routing strategy over multiple stereotype-relevant experts is designed to learn unbiased representations against different stereotypes in \llmrec{}.
    
    \item Extensive experiments on different real-world recommendation datasets are conducted to demonstrate the effectiveness of our proposed methods under various fairness settings.
\end{itemize}
\section{RELATED WORK}
In this section, we first discuss the existence and cause of intrinsic stereotypes in LLMs, then move to the existing research on the fairness issues in LLM-based recommendations.

\subsection{Stereotype of LLMs}
Stereotypes refer to social bias and discrimination that associate diverse combinations of characteristics with specific groups~\cite{kotek2023gender}, such as a combination of ``\emph{female nurse}" or ``\emph{male doctor}".
Existing studies have revealed that LLMs trained on large-scale uncurated data, particularly uncensored and imbalanced text corpora, inherit various stereotypes against specific social groups, which can be categorized by age, gender, and religion, etc.~\cite{navigli2023biases,gallegos2024bias}.
More specifically, stereotypes are ubiquitously encapsulated into the LLM vocabulary of word embeddings, leading to stereotypical behavior in various LLM-based applications since the fixed vocabulary is used to represent downstream tasks as textual information (e.g., task-specific prompts)~\cite{guo2022auto}.

Compared to conventional recommendation models, LLMs introduce unique stereotypes for inferring user preferences, leading to substantial discrimination in personalized recommendations.
In other words, discriminative predictions can be elicited by prompting LLMs with textual information of users and items. 
For example, a female stereotype can be probed by LLMs given an interacted item \emph{``dress"} in e-commerce recommendations.
Notably, recent studies have indicated that most existing LLMs, such as T5, GPT, and LLaMA families, suffer from discriminative behaviors against stereotypes (e.g., gender) in recommendation tasks~\cite{hua2023up5, zhang2023chatgpt, xu2023llms}.

\subsection{Fairness in LLM-based Recommendation}
Compared to unique user/item representations in conventional recommender systems, users and items share the textual representation (e.g., word embeddings) in LLM-based recommendations.
Recent studies have revealed that LLMs are likely to inherit stereotypes that are embedded ubiquitously in word embeddings, due to their training on large-scale uncurated datasets~\cite{navigli2023biases, gallegos2024bias}.
For example, \llmrec{} are likely to stereotypically suggest ``\emph{female nurse}" and ``\emph{male doctor}" in job recommendations~\cite{gallegos2024bias, kotek2023gender}.
This leads to \llmrec{} exhibiting stereotypical linguistic associations between users and items, causing a form of two-sided (i.e., user-to-item) recommendation fairness.
However, most existing studies on \llmrec{} fairness are limited to either user-side or item-side fairness. These technical solutions involve augmenting imbalanced datasets or fine-tuning models with fair objectives, aiming to achieve consistent recommendation performance across user groups~\cite{zhang2023chatgpt, deldjoo2024cfairllm} or fair exposure opportunities across item groups~\cite{jiang2024item, bao2024large}.
Therefore, current fairness methods might fall short in addressing two-sided recommendation fairness, since LLM stereotypes can simultaneously involve user and item groups.

In addition to the aforementioned limitation on single-sided fairness, some existing works merely leverage discrete IDs or ID embeddings of users and items for recommendations~\cite{hua2023up5, wang2024intersectional}, which implicitly ignore stereotypes in textual information, such as user profiles and item descriptions.
Notably, recent studies have indicated that textual knowledge is critical for harnessing the linguistic capabilities of \llmrec{} to effectively comprehend user preferences in recommendations~\cite{liao2023llara, zhu2024collaborative}.
To sum up, effective methods to tackle the recommendation fairness against intrinsic stereotypes of LLMs, specifically against biased textual knowledge, remain underexplored.
\section{PRELIMINARY}

In this section, we elaborate on the fairness definitions for quantifying LLM stereotypes in recommendation tasks.

\subsection{Fairness Definition}
There exist various definitions of recommendation fairness, such as consistent, calibrated, and counterfactual fairness, where the choice of these definitions depends on the scope of applications~\cite{wang2023survey, chen2023bias}.
For instance, in job recommendations, consistent fairness focuses on whether like cases are treated alike, such as whites potentially receiving more recommendations than blacks despite having equal qualifications.
In this paper, we investigate the unique characteristics of LLM
stereotypes, causing a form of two-sided (i.e., user-to-item) recommendation fairness.

To address fair recommendations, as illustrated in Fig.~\ref{fig:fairness}, we argue that user-to-item stereotypes should \textbf{not} amplify a user's preference for items.
Therefore, we extend the concept of calibrated fairness, requiring a \textbf{calibrated proportion} of stereotype groups between the original distribution (i.e., user-item interactions) and recommendation results.
Formally, given any stereotype group $G$, \fairness{} requires:
\begin{equation}
   \mathbb{P}(\mathcal{V}_u^{\bm{\mathrm{interact}}} \in G) = \mathbb{P}(\mathcal{V}_u^{\bm{\mathrm{recommend}}} \in G) \quad \forall u \in \mathcal{U},
\end{equation}
where $\mathcal{V}_u^{\bm{\mathrm{interact}}}$ and $\mathcal{V}_u^{\bm{\mathrm{recommend}}}$ are the item sets in a user $u$'s historical interactions and recommendation results, respectively. $\mathcal{U}$ denotes the user set and $\mathbb{P}$ stands for the proportion.

\subsection{Recommendation Task}

To adapt generative LLMs to recommendation tasks, recent advances have demonstrated the necessity to present target item candidates into prompts~\cite{bao2023tallrec, liao2023llara} or additional tokens~\cite{zhu2024collaborative, qu2024tokenrec} of \llmrec{}.
Therefore, we formulate recommendations as a binary classification task, where \llmrec{} will determine whether or not to recommend a given target item $v$ based on the sequence of a user $u$'s historical interactions $\mathcal{H}_u$.
\section{STEREOTYPE-AWARE FAIRNESS}\label{sec:fairness evaluation}

In pursuit of fair recommendations, we extend the concept of calibrated fairness, requiring a \textbf{calibrated proportion} of stereotype groups between the original distribution (i.e., user-item interactions) and recommendation results.

\subsection{Stereotype Group of User and Item}\label{sec:group}

As shown in Fig.~\ref{fig:group}, LLM stereotypes can simultaneously involve user and item groups. For example, a ``\emph{female}" user and an item ``\emph{dress}" can be grouped into the same gender stereotype.
Formally, let $G \in \mathcal{G}$ denote each stereotype group in a recommendation dataset, the \textbf{user-side stereotype} that $u \in G$ can be directly determined by user attributes, such as gender and age.
As for \textbf{item-side stereotype}, we interpret it as a degree $d_{v \in G}$ to which this item is mostly interacted by users of a certain stereotype group.
For example, in the case of binary stereotype groups $\mathcal{G} = [G_1, G_2]$ where an item $v$ is interacted by $30\%$ users $u \in G_1$ and $10\%$ users $u \in G_2$, the degree $d_{v \in G_1} = 0.2$ can be calculated by their subtraction.
Therefore, for any stereotype group $G \in \mathcal{G}$ (where $G' \neq G$), the degree to which an item $v$ is biased to a certain stereotype group $G$ can be calculated by
\begin{equation}\label{eq:degree}
    d_{v \in G} = \mathop{\max}_G(\frac{\sum_{u \in G}{\mathbb{1}(v \in \mathcal{H}_u)}}{\sum_{u \in G} 1} - \sum_{G' \in \mathcal{G}}{\frac{\sum_{u \in G'}{\mathbb{1}(v \in \mathcal{H}_{u'})}}{\sum_{u \in G'} 1}}),
\end{equation}
where $\mathbb{1}(v \in \mathcal{H}_u)$ equals $1$ when item $v$ is in the user $u$'s interaction history $\mathcal{H}_u$, and $0$ otherwise.

\subsection{Stereotype Measurement}

To measure the distribution of stereotype groups, given each interacted item $v$ in a user $u$'s interaction history $\mathcal{H}_u$, the proportion of any group $G$ can be formulated as:
\begin{equation}\label{eq:preference}
    h_{u \in G} = \frac{1}{|\mathcal{H}_u|}\sum_{v \in \mathcal{H}_u}{\mathbb{1}(d_{v \in G})},
\end{equation}
where $\mathbb{1}$ represents an identity function 
\begin{equation}\label{eq:threshold}
    \mathbb{1}(d_{v \in G}) = 
    \begin{cases}
        1,  & \text{if} \ d_{v \in G} \geq \mathrm{threshold} \\
        0, & \text{otherwise}
    \end{cases}
    .
\end{equation}
In practice, a threshold should be applied to the above stereotype degree $d_{v \in G}$. This is because a small degree indicates a weak bias towards any stereotype group.
For example, in Fig.~\ref{fig:fairness}, a \emph{``smartphone''} is neutral to gender stereotypes.

\begin{figure}[htbp]
\centering
{\includegraphics[width=\linewidth]{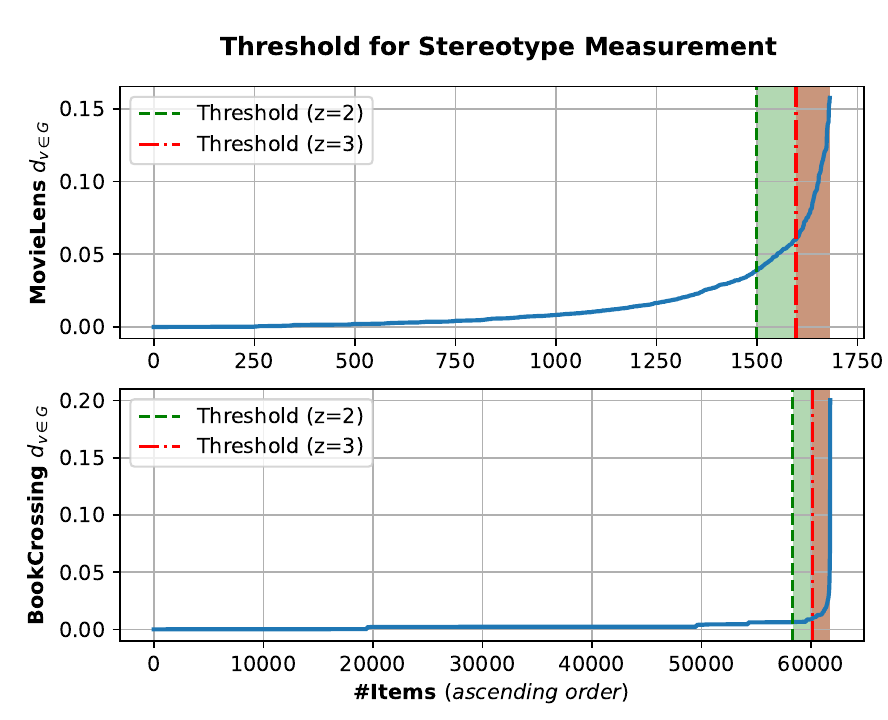}}
\caption{Threshold of $d_{v \in G}$ based on Z-scores in different experimental datasets. An ablation study is conducted to validate the above settings of threshold in Fig.~\ref{fig:ablation_item}.}
\vskip -0.15in
\label{fig:z-score}
\end{figure}

To determine the proper threshold, we employ the Z-score of $d_{v \in G}$ over all items in a recommendation dataset. 
Intuitively, Z-score identifies outliers based on how many standard deviations a data point is from the mean, where we regard outliers as items strongly related to certain stereotypes.
As illustrated in Figure~\ref{fig:z-score}, we take the commonly-used settings $z=2$ and $z=3$~\cite{vuyyala2023crop} to determine the threshold in Eq.~\eqref{eq:threshold} tailored to different experimental datasets.

\begin{table*}[htbp]
\centering
\caption{Results of preliminary experiment. In the preliminary setup, LLM-RS are fine-tuned on recommendation datasets. The reported results are multiplied by 100, where \textbf{boldface} indicates the best score.}
\label{tab:preliminary}
{
\begin{threeparttable}
\begin{tabular}{c|c|cc|ccc|ccc}
    \toprule
    \multirow{2}[2]{*}{Dataset} & \textbf{Stereotype Groups} & \multicolumn{2}{c}{\textbf{Recommendations}} & \multicolumn{3}{c}{\textbf{Fairness} $\bm{SF} \downarrow$} & \multicolumn{3}{c}{\textbf{Performance} $\uparrow$} \\
    
    & $G \in \mathcal{G}$ & users $u$ & target items $v$ & implicit & explicit & counterfactual & AUC & Precision & Recall \\
    
    \midrule
    \multirow{2}{*}{MovieLens} & \multirow{2}{*}{$\mathcal{G} = [\mathrm{male}, \mathrm{female}]$*}  & $\in G$ & $\notin G$ & \textbf{12.02} & \textbf{17.14} & \textbf{10.76} & 46.93 & 60.08 & 64.84 \\
    
    & & $\in G$ & $\in G$ & 78.82 & 83.13 & 76.11 & \textbf{72.36} & \textbf{78.25} & \textbf{76.50} \\

    \midrule
    \midrule
    \multirow{2}{*}{BookCrossing} & \multirow{2}{*}{$\mathcal{G} = [\mathrm{teen}, \mathrm{adult}]$*}  & $\in G$ & $\notin G$ & \textbf{16.08} & \textbf{25.34} & \textbf{14.56} & 71.42 & 73.09 & \textbf{61.67} \\
    
    & & $\in G$ & $\in G$ & 63.10 & 67.15 & 69.19 & \textbf{89.80} & \textbf{95.11} & 70.86 \\
    
    \bottomrule
\end{tabular}
\begin{tablenotes}
    \footnotesize
    \item * As detailed in Section~\ref{sec:setup}, available gender and age information are provided by MovieLens and BookCrossing datasets, respectively.
    
\end{tablenotes}
\end{threeparttable}
}
\vskip -0.15in
\end{table*}

\subsection{Evaluation of \FAIRNESS{}}
Based on the above stereotype measurement, extending the concept of calibrated fairness, we propose \fairness{} at the group level.
In particular, given each user in any stereotype group $G$, the recommendation proportion of target items in the same group $G$ should be calibrated to the \textbf{proportion of $\mathbf{G}$} (i.e., $h_{u \in G}$) in the user's historical interactions.
Formally, given the set of $N$ recommendations $\mathcal{S} = \{u_i, v_i\}_{i=1}^N$ where each tuple denotes the target item $v$ (or $v_i$) recommended to a user $u$ (or $u_i$), the evaluation metric of \fairness{} is defined as:
\begin{equation}\label{eq:fairness}
   \bm{SF} = 1 - \frac{1}{|\mathcal{G}|}\sum_{G \in \mathcal{G}}{\frac{\sum_{u \in \mathcal{S}}{h_{u \in G}}}{\sum_{v \in \mathcal{S}}{\mathbb{1}(d_{v \in G})}}},
\end{equation}
where $\bm{SF} > 0$ implies that \llmrec{} amplify the recommendation proportion of any stereotype group $G$ compared to that (i.e., proportion of $G$) of user-item interactions, as marked by over-recommendation in Figure~\ref{fig:fairness}.
Similarly, $\bm{SF} < 0$ indicates under-recommendation.

\section{ANALYSIS OF STEREOTYPE-AWARE FAIRNESS}\label{sec:analysis}

In this section, we initiate a preliminary experiment to investigate the influence of \fairness{}, in addressing the following two research questions:
\begin{itemize}[leftmargin=*]
    \item \textbf{RQ1:} Does \llmrec{} exhibit unfairness between the same ($u, v \in G$) and different ($u \in G, v \notin G$) stereotype groups?
    \item \textbf{RQ2:} How is \fairness{} affected by different levels of stereotypes (i.e., implicit/explicit/counterfactual)?
\end{itemize}

\subsection{Preliminary Experimental Setup}

To validate the existence of the proposed stereotype-aware fairness in LLM-RS, exploratory experiments are conducted on two widely-used recommendation datasets: MovieLens and BookCrossing.
The detailed description of datasets and evaluation metrics can be found in the main experiment section (please refer to Section~\ref{sec:setup}).

\subsection{Analysis of RQ1}

As illustrated in Table~\ref{tab:preliminary}, we compare the performance and fairness of \llmrec{} when performing recommendations between users and items from different stereotype groups.
As for performance comparisons, the recommendation quality between users and items from inconsistent stereotype groups is significantly inferior to that of consistent stereotype groups.
In particular, the AUC, Precision, and Recall of \llmrec{} decrease by 36\%, 23\%, and 15\%, respectively, in recommendations between users $u \in G$ and target items $v \notin G$, such as recommending romantic movies (e.g., female stereotype) to male users.
These observations imply that the recommendation quality of \llmrec{} is sensitive to stereotypes, emphasizing the concern of \fairness{} toward the trustworthiness of LLM-based recommendations.

Despite the downgrade in performance, the fairness of \llmrec{} regarding recommendations between inconsistent stereotype groups indicates a notably 62\%-86\% smaller value of $\bm{SF}$.
In other words, the recommendation proportion of items from a stereotype group $G$ is much more calibrated to the proportion of $G$ in user-item interactions.
However, negative $\bm{SF}$ can be observed in the MovieLens dataset, meaning that \llmrec{} rarely recommend items $v \in G$ to users $u \notin G$ despite a large proportion of $G$ in the user's historical interactions.
The aforementioned differences in recommendations between consistent and inconsistent groups imply that \llmrec{} tend to amplify the discrimination between user-side stereotypes and item-side stereotypes, such as exhibiting over-recommendation between users and items from the same stereotype group.

\subsection{Analysis of RQ2}

Since stereotypes are simultaneously embedded in the word embeddings of users and items (e.g., user profiles and item titles), we aim to investigate the degree of \fairness{} under different levels of stereotypes, namely implicit, explicit, and counterfactual settings~\cite{xu2023llms}.
Specifically, the implicit setting only provides item titles in the input prompt of \llmrec{} to infer the user-side stereotype without the actual user profile. In explicit and counterfactual settings, both user profiles and item titles are provided, as detailed in Section~\ref{sec:implementation}.
By comparing results between implicit and explicit settings, the unfairness of \llmrec{} with implicit stereotypes decreases by 29\%-36\% for inconsistent stereotype groups and 5\%-6\% for consistent stereotype groups.
These results support our findings that stereotypes exist in the word embeddings of both users and items, leading to \fairness{} in \llmrec{}.
Notably, \llmrec{} shows a reduction in unfairness by a percentage of 2\%-44\% in a counterfactual world of explicit stereotypes, highlighting the potential of utilizing different stereotypes (e.g., counterfactual stereotype) to develop effective methods in addressing \fairness{} in \llmrec{}. 

\begin{table}[htbp]
\centering
\caption{Comparison of pre-trained LLM and \llmrec{} fine-tuned on the MovieLens dataset. The reported results are multiplied by 100, where \textbf{boldface} indicates the best score.}
\label{tab:llm}
\scalebox{0.9}
{
\begin{threeparttable}
\begin{tabular}{l|ccc|ccc}
    \toprule
    \multirow{2}[2]{*}{Model} & \multicolumn{3}{c}{\textbf{Fairness} $\bm{SF} \downarrow$} & \multicolumn{3}{c}{\textbf{Performance} $\uparrow$} \\
    & implicit & explicit  & counterfactual  & AUC & Precision & Recall \\
          
    \midrule
    LLM & \textbf{60.43} & \textbf{66.87} & \textbf{57.54} & 60.96 & 65.56 & 60.47 \\
    
    \llmrec{} & 75.22 & 79.17 & 72.09 & \textbf{67.16} & \textbf{75.83} & \textbf{69.62} \\
    
\bottomrule
\end{tabular}
\end{threeparttable}
}
\vskip -0.15in
\end{table}

\subsection{Ablation on LLM-encoded Stereotype w/wo Fine-tuning.}
LLMs trained on large-scale uncurated data inherit stereotypes that are embedded ubiquitously in word embeddings. This leads to LLM-RS exhibiting stereotypical linguistic associations between users and items (e.g., user profiles and item titles).
By fine-tuning LLMs on recommendation data (i.e., \llmrec{}), we aim to investigate the influence of recommendation data to LLM-encoded stereotypes.
As illustrated in Table~\ref{tab:llm}, the performance of \llmrec{} exceeds LLM by 11\%, 15\%, and 15\% in terms of AUC, Precision, and Recall, respectively, indicating the effectiveness of fine-tuning LLMs on recommendation data.
As for the influence on stereotypes measured by \fairness{}, a significant downgrade of fairness can be observed, varying from 25\% to 39\% under different fairness settings.
This implies that the stereotypes in word embeddings of \llmrec{} can be amplified by fine-tuning on recommendation data, emphasizing the concern of \fairness{} toward the trustworthiness of LLM-RS.
\section{THE PROPOSED METHOD}

In this section, we propose a novel framework named \MOS{} along with effective learning objectives, to address \fairness{} in \llmrec{}.

\begin{figure*}[t]
    \centering
    \includegraphics[width=\textwidth]{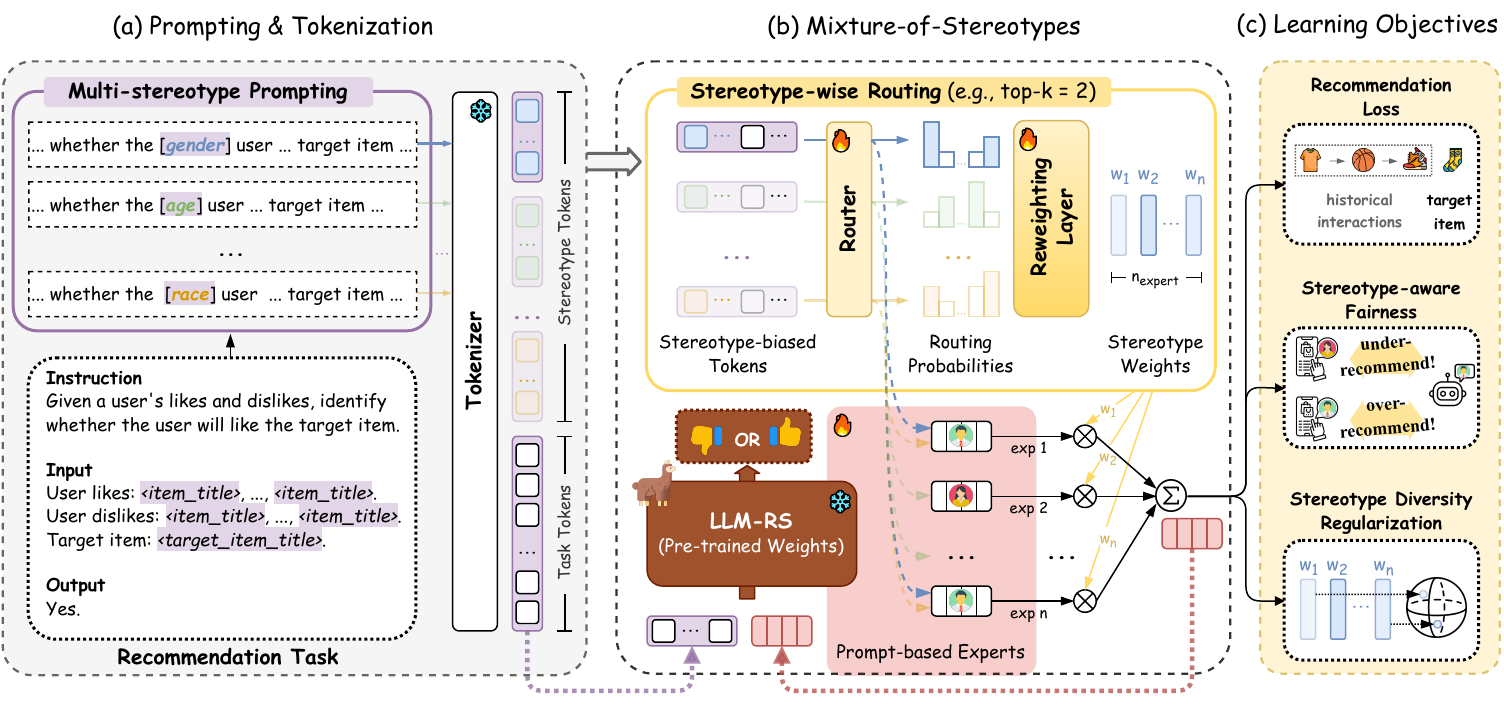}
    \caption{The overall framework of the proposed \mos{}. In (a), multi-stereotype prompting elicits biases with respect to different stereotype groups. In (b), \mos{} mitigates the elicited stereotypes in recommendation tasks, where unbiased representations are generated and integrated with \llmrec{} via soft prompts. In (c), effective learning objectives are designed to facilitate both the recommendation performance and the \fairness{}.} 
\vskip -0.15in
\label{fig:method}  
\end{figure*}

\subsection{Overview of \MOS{}}

As shown in Figure~\ref{fig:method}, our framework consists of three key modules, namely \textbf{multi-stereotype prompting} and \textbf{stereotype-wise routing}, along with effective \textbf{learning objectives}.
\begin{itemize}[leftmargin=*]
    \item 
    The first module aims to capture the representation of stereotypes in textual information.
    Since stereotypes are ubiquitously encoded in recommendation prompts, spanning over word embeddings, it is necessary to effectively distinguish different stereotype groups in addressing \fairness{}.
    Accordingly, we introduce a multi-stereotype prompting module to elicit each stereotype group via textual prompts, as recent studies have revealed that LLM stereotypes can be modified by prompts~\cite{guo2022auto, yang2023adept, li2024your}.
    
    \item In the second module, we propose a novel stereotype-wise routing strategy over a set of stereotype-relevant experts, to learn unbiased representations.
    The insight behind lies in first encoding different stereotypes into corresponding experts, then learning adaptive expert weights.
    Intuitively, the weight distribution is ``calibrated" with the original distribution of stereotype groups in user-item interactions.
    
    \item In the last module, learning objectives are carefully designed to facilitate recommendation performance and \fairness{} of \llmrec{}.
    Notebly, we introduce a stereotype diversity regulation term, to enhance the learning of stereotype information via multiple expert networks.
    
\end{itemize}

\subsection{Multi-stereotype Prompting}

As illustrated in Figure~\ref{fig:method}, a multi-stereotype prompting module is applied to the textual input prompt of recommendation tasks.
The idea of multi-stereotype prompting is to amplify the LLM-encoded stereotypes in recommendation tasks, such as gender, age, and race.
In particular, we design stereotype-biased prompts by inserting textual descriptions with respect to each stereotype group, as recent studies have revealed that stereotypes in LLMs can be modified by prompts~\cite{guo2022auto, yang2023adept, li2024your}.
For example, a prompting template for gender stereotypes is demonstrated as follows:
\begin{tcolorbox}[colback=white]
    \textbf{[Instruction]}
    Given a \colorbox{violet!15}{\emph{female}} user's interaction history, identify whether this \colorbox{violet!15}{\emph{female}} user will like the item.
\end{tcolorbox}
\noindent Formally, given a recommendation task prompt $x^{\mathrm{rec}}$ and a set of stereotype-biased prompts $\{x_i^{\mathrm{stereotype}}\}_{i=1}^{|\mathcal{G}|}$ (e.g., ``\colorbox{violet!15}{\emph{female}}"), the tokenization output of multi-stereotype prompting is formalized as
\begin{equation}
    x^{\mathrm{rec}}, \{x_i^{\mathrm{stereotype}}\}_{i=1}^{|\mathcal{G}|} \rightarrow \{(c^{\mathrm{rec}}, c_i^{\mathrm{stereotype}})\}_{i=1}^{|\mathcal{G}|},
\end{equation}
where each set of stereotype tokens $c_i^{\mathrm{stereotype}}$ are concatenated to recommendation task tokens $c^{\mathrm{rec}}$.
It is worth noting that stereotype tokens will not be applied to the input of \llmrec{}, without requiring the actual user profile in the inference stage of recommendations.

\subsection{Stereotype-wise Routing}

Following the multi-stereotype prompting module, a set of input tokens can be generated with respect to each stereotype group $G \in \mathcal{G}$.
Subsequently, a stereotype-wise routing module is developed to learn unbiased representations against stereotypes embedded in the input tokens.
Overall, the stereotype-wise routing module is composed of three key components as follows.

\subsubsection{Router}
The router aims to capture different stereotype information of each group by learning the routing strategy of forwarding stereotype-specific tokens to stereotype-relevant experts, where each expert specializes in a particular subset of stereotypes.
Given each set of input tokens $\bm{c}_i=(c^{\mathrm{rec}}, c_i^{\mathrm{stereotype}})$, the router $\bm{Q}$ determines the routing probabilities to $N$ experts as follows:
\begin{equation}
    \{p_n(\bm{c}_i)\}_{n=1}^N = \bm{Q}(\bm{c}_i),
\end{equation}
where $p_n$ denotes the probability of forwarding input tokens to the $n$-th expert.
In other words, the router is trained to assign different stereotype groups in \llmrec{} to multiple experts. 

Subsequently, the stereotype-specific tokens can be forwarded to corresponding stereotype-relevant experts based on the top-K routing strategy.
In particular, given the top-K highest probabilities determined by the router, the modified routing probability of each $n$-th expert can be obtained by
\begin{equation}\label{eq:topk}
    p_n(\bm{c}_i) = 
    \begin{cases}
        [softmax(\{p_k(\bm{c}_i)\}_{k \in \mathcal{K}})]_n,  & \text{if} \ n = k \\
        0, & \text{otherwise}
    \end{cases}
    ,
\end{equation}
where $k \in \mathcal{K}$ denotes the index of each activated expert in the top-K set $\mathcal{K}$ (i.e., $|\mathcal{K}| \leq N$).
It is worth noting that the routing probabilities of inactivated experts are set to zero, since a reweighting strategy of the routing probabilities will be designed, as illustrated in Section~\ref{sec:reweight}.
Intuitively, the zeroing operation reinforces the polarization of routing different stereotype information to different experts, facilitating the goal of stereotype-wise routing.

\subsubsection{Reweighting Layer}\label{sec:reweight}

In light of our preliminary findings that the discrimination between user-side and item-side stereotypes can be alleviated in a counterfactual world, we explore combining multiple stereotypes (i.e., stereotype-relevant experts) to address \fairness{} in \llmrec{}.
In particular, we aim to generate unbiased representations that are calibrated with the original distribution of stereotypes in user-item interactions, by learning adaptive weights across different stereotype-relevant experts.
As a natural solution, the learning objectives of adaptive weights can be obtained by \fairness{}, as illustrated in Eq.~\eqref{eq:fairness}, and added to the training loss of \llmrec{}.
However, unlike discriminative recommendation models, it is challenging to update generative \llmrec{} with a learning objective given by group-level fairness~\cite{dai2024bias, atwood2024inducing, tommasel2024fairness}.
To be specific, the learning objectives of generative \llmrec{} are to maximize the likelihood of the label tokens of each output (e.g., target item), which intrinsically fall short in calculating an auxiliary loss over a group of outputs.

To address these challenges, a reweighting layer is designed to pre-calculate the adaptive weights across stereotype-relevant experts based on their routing probabilities.
Formally, let $\bm{R}$ denote the reweighting layer, the stereotype weights of each expert can be calculated by
\begin{equation}
    \{w_n\}_{n=1}^N = \bm{R}(\frac{1}{|\mathcal{G}|}\sum_{i=1}^{|\mathcal{G}|}{\{p_n(\bm{c}_i)\}_{n=1}^N}).
\end{equation}
In other words, $\{w_n\}_{n=1}^N$ implies a weighted average of multiple stereotypes encoded in each corresponding stereotype-relevant expert.

\subsubsection{Prompt-based Experts}

With the aforementioned reweighting strategy, the mixture of stereotype-relevant experts can generate unbiased representations against stereotypes, taking advantage of the weighted average of different stereotypes that accord with the original distribution of stereotypes in user-item interactions.
To adapt the learned unbiased representations to \llmrec{}, we design prompt-based experts to encode the generated representations as soft prompts, which can be seamlessly integrated with \llmrec{} via prompt tuning~\cite{lester2021power}.
Formally, the soft prompts generated by experts $\{\bm{E}_n\}_{n=1}^N$ can be formalized as:
\begin{equation}
    e = \sum_{n=1}^N{w_n\bm{E}_n(c^\mathrm{rec})}.
\end{equation}
Thereafter, the final recommendation output of \llmrec{} with the proposed \mos{} module is as follows:
\begin{equation}
    y = \text{\llmrec{}}(e, c^\mathrm{rec}),
\end{equation}
where the pre-trained weights of \llmrec{} are frozen.

\subsection{Learning Objectives}

Let $\Phi$ denote the frozen pre-trained weights of LLMs and $\Theta$ be the learnable parameters of \mos{} (i.e., router, reweighting layer, and experts), the learning objectives consist three terms, namely \textbf{recommendation performance}, \textbf{\fairness{}}, and \textbf{stereotype diversity regularization}.
In particular, the recommendation loss of generative \llmrec{} is given by
\begin{equation}
    \mathcal{L}_\mathrm{rec} = - \log \Pr_{\Phi + \Theta}(\hat{y} | e, c^\mathrm{rec}),
\end{equation}
where $\hat{y}$ denotes the label tokens of recommendation outputs.
As for the fairness loss, we apply the proposed evaluation metric of \fairness{}, which can be defined as: 
\begin{equation}
\mathcal{L}_\mathrm{fair} \coloneq \mathop{\min}_\Theta \Vert \bm{SF} \Vert.
\end{equation}
Notably, we further introduce a stereotype diversity regulation term to enhance the learning of stereotype information via multiple expert networks.
In detail, expert parameter redundancy is invertible in multiple-expert architectures, leading to similar representations (i.e., learnable knowledge) across multiple expert networks~\cite{liu2023fair, tian2024dialogue, dai2024deepseekmoe}.
Therefore, a stereotype diversity regulation term is designed to maximize the distance between the weights of each stereotype-relevant expert (i.e., diversified representations) as follows:
\begin{equation}
    \mathop{\max}_{\{w_n\}_{n=1}^N}\{\mathcal{L}_\mathrm{expert}
    \coloneq \mathop{\min}_{i \neq j}(\Vert w_i - w_j \Vert^2)\}.
\end{equation}
Finally, the overall learning objectives of \llmrec{} with the proposed \mos{} module can be formalized as: 
\begin{equation}\label{eq:loss}
\mathcal{L}_\mathrm{total} = \mathcal{L}_\mathrm{rec} + \mathcal{L}_\mathrm{fair} + \mathcal{L}_\mathrm{expert}.
\end{equation}
\section{EXPERIMENT}

Following our preliminary findings, extensive experiments are conducted to demonstrate the superiority of proposed methods under various fairness settings of \llmrec{}.

\subsection{Experiment Setup}\label{sec:setup}

\subsubsection{Datasets}
We conducted experiments on two datasets, which contain user profiles regarding different stereotypes. 

\noindent\textbf{MovieLens100K}~\cite{harper2015movielens} is a movie recommendation dataset, which provides user-movie interactions and textual information including movie titles and user profiles. In particular, we utilize the binary gender feature of users to assess the gender stereotype of \llmrec{}.

\noindent\textbf{BookCrossing}~\cite{ziegler2005improving} is a book recommendation dataset, which provides user-book interactions and textual information including book titles and user profiles. In particular, we utilize the age feature and divide users into teen and adult groups (i.e., under/beyond 18), to assess the age stereotype of \llmrec{}.

To maintain a manageable dataset size for efficient \llmrec{} training, similar to recent studies~\cite{liao2023llara, bao2023tallrec}, we process the original datasets by randomly sampling 10,000 sequences (i.e., each contains 11 chronological interactions).
To construct sequential recommendation scenarios, we adopt the leave-one-out strategy and retain the first 10 items in each sequence as the historical interaction, and the last item as the target item.
For both datasets, we split the data points of user-item interactions into training, validation, and testing sets with a ratio of 8:1:1, which prevents data leakage.
The detailed statistics of experimental datasets are provided in Table~\ref{tab:dataset}.
\begin{table}[htbp]
\centering
\caption{Basic statistics of experimental datasets.}
\label{tab:dataset}
\scalebox{0.95}
{
\begin{threeparttable}
\begin{tabular}{c|c|c|c|c}

\toprule
\multirow{2}{*}{\textbf{Datasets}} & \multicolumn{3}{c}{\small\textbf{User-Item Interaction}} & \small\textbf{Group Ratio} \\

& \small\#Users & \small\#Items & \small\#Interactions & \small$G_1:G_2$*\\ 

\midrule
\small\textbf{MovieLens} & 943  & 1,682 & 19,688 & $\approx$7:3\\

\small\textbf{BookCrossing} & 62,649 & 61,740 & 23,238 & $\approx$6:4\\

\bottomrule
\end{tabular}
\begin{tablenotes}
\footnotesize
    \item $*:$ The pair $G_1$/$G_2$ denotes the male/female stereotype group pair and the adult/teen stereotype group pair, respectively, in MovieLens and BookCrossing.
\end{tablenotes}
\end{threeparttable}
}
\vskip -0.15in
\end{table}

\subsubsection{Evaluation Metrics}\label{sec:metrics}

We assess the proposed \fairness{} and the recommendation performance.

\noindent\textbf{Fairness.} The proposed \fairness{} $\bm{SF}$ can be calculated according to Eq.~\eqref{eq:fairness}. A smaller value of $\bm{SF}$ suggests a more calibrated (i.e., fair) proportion of stereotype groups between recommendation results and the original distribution in user-item interactions.

\begin{table*}[htbp]
  \centering
  \caption{Results of fairness-oriented methods. We report the average results over three independent runs with random seeds. The reported results are multiplied by 100, where \textbf{boldface} and \underline{underline} indicate the best and second best score, respectively. The {\color{ForestGreen}{positive}} and {\color{Maroon}{negative}} improvements (\%) of our proposed method are compared to the best baseline.}
  \label{tab:fairness}
{
\begin{threeparttable}
\begin{tabular}{l|l|c|ccc|ccc}
    \toprule
    \multirow{2}[2]{*}{Dataset} & \multirow{2}[2]{*}{Method*} & \multirow{2}[2]{*}{Stakeholder} & \multicolumn{3}{c}{\textbf{Fairness} $\bm{SF} \downarrow$} & \multicolumn{3}{c}{\textbf{Performance} $\uparrow$} \\
    & & & implicit & explicit & counterfactual & AUC & Precision & Recall \\
    
    \midrule
    \multirow{4}{*}{MovieLens} & IFairLRS & item-side & 83.78 & 78.20 & 75.06 & \underline{69.20} & \underline{77.33} & \textbf{55.95} \\

    & UP5 & user-side & 75.31 & 72.48 & 74.12 & 67.60 & 75.40 & \underline{52.73} \\

    & FaiRLLM & user-side & \underline{63.18} & \underline{67.40} & \underline{68.41} & 66.77 & 74.87 & 47.91 \\

    & \cellcolor{gray!15}{\textbf{\mos{}}} & \cellcolor{gray!15}{two-sided} & \cellcolor{gray!15}{\textbf{55.49}({\color{ForestGreen}{12.2\%}})} & \cellcolor{gray!15}{\textbf{53.71}({\color{ForestGreen}{20.3\%}})} & \cellcolor{gray!15}{\textbf{50.91}({\color{ForestGreen}{25.6\%}})} & \cellcolor{gray!15}{\textbf{69.64}({\color{ForestGreen}{0.6\%}})} & \cellcolor{gray!15}{\textbf{77.83}({\color{ForestGreen}{0.6\%}})} & \cellcolor{gray!15}{50.80({\color{Maroon}{9.2\%}})} \\
    
    \midrule
    \midrule
    \multirow{4}{*}{BookCrossing} & IFairLRS & item-side & 74.10 & 67.28 & 70.65 & \underline{78.32} & 70.83 & \underline{59.38} \\

    & UP5 & user-side & \underline{58.67} & \underline{60.00} & \underline{59.38} & 78.71 & \textbf{87.53} & 55.52 \\

    & FaiRLLM & user-side & 65.32 & 66.67 & 62.28 & 58.46 & 66.52 & 50.41 \\

    & \cellcolor{gray!15}{\textbf{\mos{}}} & \cellcolor{gray!15}{two-sided} & \cellcolor{gray!15}{\textbf{55.71}({\color{ForestGreen}{5.0\%}})} & \cellcolor{gray!15}{\textbf{50.04}({\color{ForestGreen}{16.6\%}})} & \cellcolor{gray!15}{\textbf{57.24}({\color{ForestGreen}{3.6\%}})} & \cellcolor{gray!15}{\textbf{79.15}({\color{ForestGreen}{1.1\%}})} & \cellcolor{gray!15}{\underline{79.39}({\color{Maroon}{9.3\%}})} & \cellcolor{gray!15}{\textbf{72.98}({\color{ForestGreen}{22.9\%}})} \\
    
    \bottomrule
\end{tabular}
\begin{tablenotes}
\footnotesize
    \item $*:$ The learning objectives $\mathcal{L}_\mathrm{fair}$ in Eq.~\eqref{eq:loss} are replaced by the corresponding fairness metrics in each baseline of fairness-oriented methods.
\end{tablenotes}
\end{threeparttable}
}
\end{table*}

\begin{table*}[htbp]
  \centering
  \caption{Results of \llmrec{} paradigms. We report the average results over three independent runs with random seeds. The reported results are multiplied by 100, where \textbf{boldface} and \underline{underline} indicate the best and second best score, respectively.}
  \label{tab:llmrs}
{
\begin{threeparttable}
\begin{tabular}{l|l|c|c|ccc|ccc}
    \toprule
    \multirow{2}[2]{*}{Dataset} & \multirow{2}[2]{*}{Method} & \multirow{2}{*}{
    \begin{tabular}[c]{@{}c@{}}
        Trainable \\
        Params (\%)
    \end{tabular}
    } & \multirow{2}{*}{
    \begin{tabular}[c]{@{}c@{}}
        Learning \\
        Objectives
    \end{tabular}
    } & \multicolumn{3}{c}{\textbf{Fairness} $\bm{SF} \downarrow$} & \multicolumn{3}{c}{\textbf{Performance} $\uparrow$} \\
    & & & & implicit & explicit & counterfactual & AUC & Precision & Recall \\
    
    \midrule

    \multirow{7}{*}{MovieLens} & Full-model tuning  & - & \multirow{2}{*}{$\mathcal{L}_{\mathrm{rec}}$} & 75.22 & 79.17 & 72.09 & 67.16 & 75.83 & 69.62 \\

    & \textbf{\mos{}} & 0.0331 & & 72.47 & 80.26 & 70.45 & 65.86 & 69.07 & 64.69 \\

    \cmidrule{2-10}

    & Full-model tuning  & - & $\mathcal{L}_\mathrm{total} *$ & \underline{62.39} & \underline{62.83} & 61.12 & \underline{68.91} & \underline{73.12} & \underline{62.20} \\
    
    & \cellcolor{gray!15}{\textbf{\mos{}} ({\color{blue}{\textbf{multi-expert}}})} & \cellcolor{gray!15}{0.0331} & \cellcolor{gray!15}{$\mathcal{L}_\mathrm{total}$} & \cellcolor{gray!15}{\textbf{55.49}} & \cellcolor{gray!15}{\textbf{53.71}} & \cellcolor{gray!15}{\textbf{50.91}} & \cellcolor{gray!15}{\textbf{69.64}} & \cellcolor{gray!15}{\textbf{77.83}} & \cellcolor{gray!15}{50.80} \\

    & \cellcolor{gray!15}{- LoRA (adapter)} & \cellcolor{gray!15}{0.3954} & \cellcolor{gray!15}{$\mathcal{L}_\mathrm{total} *$} & \cellcolor{gray!15}{63.25} & \cellcolor{gray!15}{64.53} & \cellcolor{gray!15}{61.27} & \cellcolor{gray!15}{68.25} & \cellcolor{gray!15}{72.55} & \cellcolor{gray!15}{\textbf{70.29}} \\

    & \cellcolor{gray!15}{- p-tuning (encoder)} & \cellcolor{gray!15}{0.1097} & \cellcolor{gray!15}{$\mathcal{L}_\mathrm{total} *$} & \cellcolor{gray!15}{67.60} & \cellcolor{gray!15}{72.07} & \cellcolor{gray!15}{\underline{60.49}} & \cellcolor{gray!15}{67.23} & \cellcolor{gray!15}{72.90} & \cellcolor{gray!15}{55.79} \\

    & \cellcolor{gray!15}{- prompt-tuning (embedding)} & \cellcolor{gray!15}{0.0138} & \cellcolor{gray!15}{$\mathcal{L}_\mathrm{total} *$} & \cellcolor{gray!15}{65.53} & \cellcolor{gray!15}{66.25} & \cellcolor{gray!15}{61.43} & \cellcolor{gray!15}{56.76} & \cellcolor{gray!15}{67.65} & \cellcolor{gray!15}{44.05} \\

    \midrule
    \midrule
    \multirow{7}{*}{BookCrossing} & Full-model tuning  & - & \multirow{2}{*}{$\mathcal{L}_{\mathrm{rec}}$} & 59.10 & 62.39 & 59.28 & 77.59 & 80.62 & 64.14 \\

    & \textbf{\mos{}} & 0.0331 & & 58.25 & 60.02 & 59.45 & 77.22 & 76.45 & 68.18 \\

    \cmidrule{2-10}

    & Full-model tuning  & - & $\mathcal{L}_\mathrm{total} *$ & 56.14 & \underline{58.67} & 60.96 & 78.41 & \underline{84.25} & 68.05 \\

    & \cellcolor{gray!15}{\textbf{\mos{} ({\color{blue}{\textbf{multi-expert}}})}} & \cellcolor{gray!15}{0.0331} & \cellcolor{gray!15}{$\mathcal{L}_\mathrm{total}$} & \cellcolor{gray!15}{\underline{55.71}} & \cellcolor{gray!15}{\textbf{50.04}} & \cellcolor{gray!15}{\textbf{57.24}} & \cellcolor{gray!15}{\textbf{79.15}} & \cellcolor{gray!15}{79.39} & \cellcolor{gray!15}{\underline{72.98}} \\

    & \cellcolor{gray!15}{- LoRA (adapter)} & \cellcolor{gray!15}{0.3954} & \cellcolor{gray!15}{$\mathcal{L}_\mathrm{total} *$} & \cellcolor{gray!15}{57.96} & \cellcolor{gray!15}{61.35} & \cellcolor{gray!15}{61.81} & \cellcolor{gray!15}{78.54} & \cellcolor{gray!15}{81.52} & \cellcolor{gray!15}{63.26} \\

    & \cellcolor{gray!15}{- p-tuning (encoder)} & \cellcolor{gray!15}{0.1097} & \cellcolor{gray!15}{$\mathcal{L}_\mathrm{total} *$} & \cellcolor{gray!15}{58.35} & \cellcolor{gray!15}{63.65} & \cellcolor{gray!15}{64.29} & \cellcolor{gray!15}{\underline{79.01}} & \cellcolor{gray!15}{77.49} & \cellcolor{gray!15}{\textbf{74.30}} \\

    & \cellcolor{gray!15}{- prompt-tuning (embedding)} & \cellcolor{gray!15}{0.0138} & \cellcolor{gray!15}{$\mathcal{L}_\mathrm{total} *$} & \cellcolor{gray!15}{\textbf{54.07}} & \cellcolor{gray!15}{60.30} & \cellcolor{gray!15}{\underline{60.56}} & \cellcolor{gray!15}{78.34} & \cellcolor{gray!15}{\textbf{88.18}} & \cellcolor{gray!15}{43.00} \\
    
    \bottomrule
    
\end{tabular}
\begin{tablenotes}
\footnotesize
    \item $*:$ To validate the effectiveness of our proposed {\color{blue}{\textbf{multi-expert}}} structure in mitigating different stereotypes, all baselines of conventional structures (e.g., LoRA) are compared using the same learning objectives. Since no expert module is involved in baselines, $\mathcal{L}_\mathrm{expert}$ is excluded in $\mathcal{L}_\mathrm{total}$.
\end{tablenotes}
\end{threeparttable}
}
\end{table*}

\noindent\textbf{Performance.}
Following the recent implementation of LLM-based recommendations~\cite{hua2023up5, bao2023tallrec, xi2024towards}, we adopt the area under the receiver operating characteristic (AUC) for performance evaluation.
Meanwhile, considering the imbalanced classes (i.e., stereotype groups) in real-world recommendation datasets, the model may frequently make incorrect predictions for minority classes, yet still achieve a high AUC.
Therefore, we further compare the precision and recall for performance evaluation.

\subsubsection{Baselines}

We design two groups of baselines to investigate the fairness improvement compared to current fairness-oriented methods and the effectiveness of the proposed \mos{} framework compared to conventional \llmrec{} paradigms.

\noindent\textbf{Fairness-oriented Methods.} IFairLRS~\cite{jiang2024item} proposes a reweighting strategy to mitigate biases stemming from unbalanced groups, where sample weights are applied to the loss of instruction-tuning samples of \llmrec{}.
UP5~\cite{hua2023up5} introduces a counterfactually fair prompting method, which masks sensitive user information by prompt tuning with a discrimination loss.
FaiRLLM~\cite{zhang2023chatgpt} designs fairness metrics
tailored to \llmrec{} by mitigating the divergence of similarity metrics of recommendations against sensitive attributes.

\noindent\textbf{\llmrec{} Paradigms.} To validate the effectiveness of our proposed multi-expert structure in mitigating different stereotypes, we compare \mos{} to conventional training paradigms of \llmrec{}.
The baselines include full-model tuning and parameter-efficient fine-tuning (PEFT) paradigms, such as prompt tuning~\cite{lester2021power}, p-tuning~\cite{liu2022p}, and LoRA~\cite{hu2022lora}.
In particular, LoRA adds extra trainable adapter weights that are merged with LLM parameters. P-tuning trains a prompt encoder to generate learnable prompt embeddings in the LLM input. Similarly, prompt-tuning directly updates the prompt embeddings.
For a fair comparison between MoS (i.e., multi-expert) and baselines (i.e., non-expert), the learnable modules of each PEFT baseline are trained using the same learning objectives.

\subsubsection{Implementation Details}\label{sec:implementation}

To implement \llmrec{} in a generative paradigm, the recommendation task is formulated into prompts.
In particular, the personalized preference of users is indicated based on median ratings, which are 3 and 5 in MovieLens and BookCrossing datasets, respectively. An example prompt is provided below:
\begin{tcolorbox}[colback=white]
    \textbf{[Instruction]}
    Given a user's interaction history, identify whether this user will like the target item.
    \tcblower
    \textbf{[Input]}
    
    User likes: $\langle item\_titles \rangle$ (rating $\geq$ median value)
    
    User dislikes: $\langle item\_titles \rangle$ (rating $<$ median value)
    
    Target item: $\langle item\_title \rangle$
\end{tcolorbox}

Our proposed methods are implemented based on HuggingFace and PyTorch.
For a fair comparison across all baselines, we employ a widely-used lightweight LLM, i.e., T5~\cite{raffel2020exploring}, with encoder-decoder structures as the backbone model of \llmrec{}.
As for the proposed \mos{} framework, we employ linear stereotype-wise routing models (i.e., router and reweighting layer) and $4$ prompt-based experts, each with a length of $5$ tokens for eliciting different stereotypes in personalized recommendations. 
In particular, we optimize the aforementioned proposed models with Adafator~\cite{shazeer2018adafactor}, where the learning rates for prompt-based baselines (i.e., p-tuning, prompt tuning, and \mos{}) and adapter-based baselines (i.e., TALLRec and LoRA) are set to be $0.5$ and $0.005$, respectively.

\subsection{Performance Comparison}

\subsubsection{Comparison of Fairness-oriented Methods}

As shown in Table~\ref{tab:fairness}, we compare the fairness and performance between our proposed \mos{} and existing fairness-oriented methods of \llmrec{}. 
Overall, \mos{} significantly outperforms the current single-sided fairness baselines (i.e., user-side and item-side) to mitigate stereotypes in \llmrec{}.
In the meanwhile, \mos{} achieves slightly better or comparable recommendation performance compared to each baseline. 

Comparing our proposed \mos{} to single-sided fairness methods of \llmrec{}, current user-side fairness methods (e.g., UP5 and FairRLLM) indeed contribute to facilitating \fairness{}, and outperform item-side fairness methods (e.g., IFairLRS) by 10\%-24\%.
These improvements persist even when utilizing more intricate stereotype settings, allowing them to partially address fairness against stereotypes.
We infer that user-side fairness methods potentially alleviate the stereotypes in word embeddings by eliminating discrimination between user groups with different stereotypes.
However, the fairness performance gap compared to our proposed \mos{} is still significant, indicating that current fairness-oriented methods lack effective mechanisms to address the recommendation biases between user-side stereotypes and item-side stereotypes.

In terms of the fairness performance against different types of stereotypes, \mos{} outperforms all baselines by 12\%-25\% for gender stereotypes in the MovieLens dataset and 3\%-16\% for age stereotypes in the BookCrossing dataset.
It is worth noting that the \fairness{} of \mos{} no longer exhibits particular patterns under different levels of stereotypes (i.e., implicit, explicit, and counterfactual settings) as illustrated in \textbf{RQ2}, implying the effectiveness to mitigate stereotypes in the wording embeddings of both users and items in \llmrec{}.

\subsubsection{\mos{} vs. Conventional LLM-RS Paradigms}

To demonstrate the effectiveness of \mos{} to learn unbiased representations utilizing multiple stereotype-relevant experts, we design baselines by replacing the \mos{} framework (i.e., multiple experts with stereotype-wise routing) with conventional PEFT methods to train \llmrec{} under various setups of learning objectives, as shown in Table~\ref{tab:llmrs}.

Comparing the fairness performance of training \llmrec{} with the proposed \mos{} framework to conventional PEFT paradigms (i.e., prompt tuning, p-tuning, and LoRA), it can be noticed that \mos{} consistently outperforms all baselines by 5\%-14\% in average fairness and achieves comparable AUC performance between 0.1\% and 18\% compared to baselines, even with higher trainable parameters of 331\%-1194\%.
Notably, significant trade-offs between fairness and performance can be noticed in some conventional \llmrec{} paradigms, while \mos{} maintains the best or second best scores of both fairness and performance in most situations.

\begin{table}[htbp]
\centering
\caption{Results of ablation studies on \mos{} components. We report the average fairness ($\bm{SF}$) and performance (AUC) over three independent runs with random seeds.}
\label{tab:components}
\begin{threeparttable}
\begin{tabular}{lcccc}
    \toprule
    \multirow{2}[2]{*}{Module} & \multicolumn{2}{c}{MovieLens} & \multicolumn{2}{c}{BookCrossing} \\
    & $\bm{SF} \downarrow$  & AUC $\uparrow$  & $\bm{SF} \downarrow$  & AUC $\uparrow$ \\
          
    \midrule
    \cellcolor{gray!15}{\textbf{MoS} (top-$1$)} & \cellcolor{gray!15}{\textbf{55.49}} & \cellcolor{gray!15}{\textbf{69.64}} & \cellcolor{gray!15}{55.71} & \cellcolor{gray!15}{\textbf{79.15}} \\
    
    - top-$2$ & 70.00  & 63.34  & 62.03  & 75.59  \\
    
    - stochastic & 82.39  & 59.80  & \textbf{31.11}  & 53.68  \\
    
    - w/o reweighting & 64.78  & 68.48  & 60.70  & 78.61  \\
    
    - w/o $\mathcal{L}_\mathrm{div}$ & 57.60  & 69.45 & 53.62  & 78.92  \\

\bottomrule

\end{tabular}
\end{threeparttable}
\vskip -0.15in
\end{table}

Delving into the trade-offs between Precision and Recall, interesting patterns can be observed between the learning objectives with and without \fairness{}.
To be specific, \llmrec{} trained with fairness loss mostly exhibit higher scores of Precision than Recall, and vice versa.
This implies a reduction of false positive predictions when recommending negative target items to users. Based on the findings in \textbf{RQ1} that \llmrec{} exhibit over-recommendations (i.e., false positive predictions) between users and items from the same stereotype, the Precision improvements indicate the effectiveness of \fairness{} to mitigate stereotypes between users and items.

\subsection{Ablation Study}

\subsubsection{\mos{} Components}

To assess the influence of each key component, we conducted ablation experiments on the effectiveness of \mos{} with separately eliminated components, as shown in Table~\ref{tab:components}.
In particular, we compare the stereotype-wise routing component between top-1, top-2, and stochastic routing strategies, as illustrated in Eq.~\eqref{eq:topk}.
Notably, the top-1 setting outperforms top-2 and stochastic settings in terms of both fairness and performance.
We speculate that the top-2 routing strategy potentially encodes different stereotype information to the same experts.
In addition to the routing strategy, we compare the effectiveness of \mos{} without the reweighting layer and corresponding learning objectives $\mathcal{L}_{div}$.
Despite comparable results in performance, the proposed components significantly improve the fairness by 3\%-16\%.

\begin{figure}[htbp]
\centering
{\includegraphics[width=\linewidth]{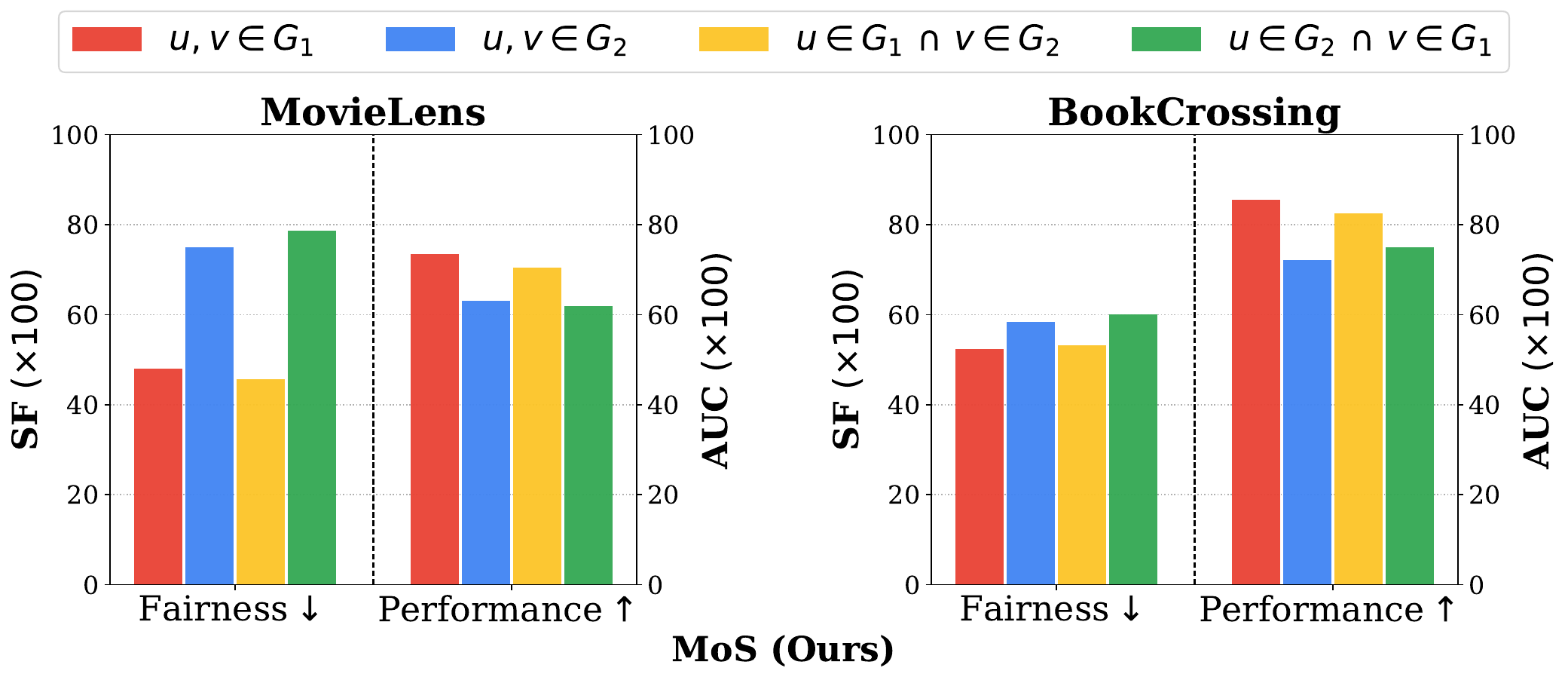}}
\caption{Results of ablation studies on user-side and item-side stereotypes. The detailed statistics of users $u$, items $v$, and groups $G_1, G_2$ can be found in Table~\ref{tab:dataset}.}
\label{fig:ablation_user}
\end{figure}

\subsubsection{User-side vs. Item-side Stereotypes}

Since a stereotype group contains both users and items in \llmrec{}, we further delve into the effectiveness of \mos{} by separately comparing user-side stereotype and item-side stereotypes, as shown in Figure~\ref{fig:ablation_user}.
By comparing between red/blue and yellow/green bars, it can be observed that \mos{} can achieve consistent fairness and performance between users and items of different stereotype groups.
However, the fairness and performance of user stereotype group $u \in G_1$ (i.e., red and yellow bars) significantly exceed that of $u \in G_2$ (i.e., blue and green bars).
One likely reason is that the amount of user-item interactions is dominated by $u \in G_1$ due to the unequal distribution of user groups in real-world recommendation datasets, as illustrated in Table~\ref{tab:dataset}, leading to more sampled training data.

\begin{figure}[htbp]
\centering
{\includegraphics[width=0.95\linewidth]{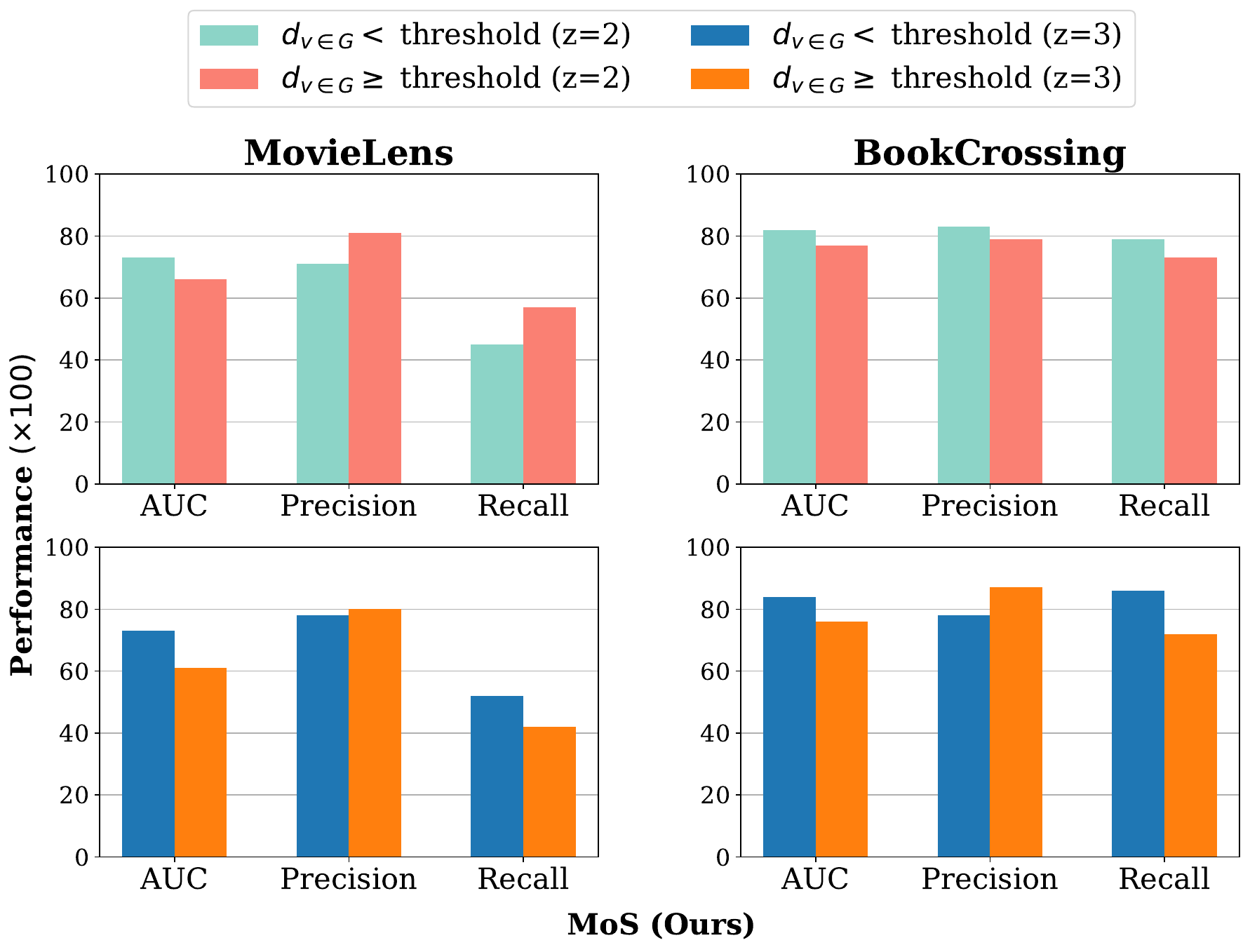}}
\caption{Results of ablation studies on the threshold of stereotype measurement based on Z-scores.}
\vskip -0.15in
\label{fig:ablation_item}
\end{figure}

\subsubsection{Threshold for Stereotype Measurement}\label{sec:z-score}
To validate the threshold of stereotype measurement based on Z-scores, we compare the recommendation performance of target items below and above the designed threshold.
As revealed by our preliminary experiments in Table~\ref{tab:preliminary}, items below the threshold are weakly biased to any stereotype group, indicating a lower degree of stereotype (i.e., measured by $d_{v \in G}$) that potentially downgrades \llmrec{} performance.
It is worth noting that the comparison of fairness metrics is ignored since these items (i.e., below the threshold) are not involved in any stereotype group.
As shown in Figure~\ref{fig:ablation_item}, the recommendation performance of items below the threshold notably exceeds that of items above the threshold, implying that Z-scores can identify an effective threshold for stereotype measurement.
\section{CONCLUSION}
This study investigates the unique characteristics of LLM stereotypes in recommendation fairness, which simultaneously involves user and item groups (i.e., two-sided groups).
In this paper, we propose a new variant of fairness, extending the concept of calibrated fairness, to evaluate the influence of LLM stereotypes on a two-sided (i.e., user-to-item) recommendation fairness.
In other words, the stereotypical linguistic associations, encoded by LLMs, between users and items should not amplify a user's preference for items in recommendation tasks.
To mitigate unfairness due to stereotypes in \llmrec{}, a novel framework called \mos{} is proposed along with effective learning objectives.
In particular, we develop multiple stereotype-relevant experts to capture different stereotypes in textual user and item information, where an insightful stereotype-wise routing strategy is designed to learn unbiased representations against different stereotypes.
Through comprehensive experiments on recommendation datasets under various fairness settings, we demonstrated the effectiveness of the proposed methods in addressing \fairness{} of \llmrec{}.
As for future work, further investigation might shed light on the individual-level fairness against diverse combinations of stereotypes in LLM-based recommendations.

\bibliographystyle{IEEEtran}
\bibliography{IEEEabrv, reference}

\vfill

\end{document}